\documentclass{article} 
\usepackage{amsmath,amsfonts,amssymb,graphicx,color, bm, xcolor, units}
\usepackage[colorlinks=true,citecolor=blue]{hyperref}
\usepackage{soul}
\usepackage{soul}
\usepackage[switch]{lineno}
\usepackage[affil-it]{authblk}
\newcommand{\de}{\partial}

\title{\textbf{Shear viscosity to entropy density ratio: A powerful tool for gravity theories and strongly coupled fluids}}

\author[1,2]{Marilù Chiofalo}
\author[2]{Dario Grasso}
\author[3,4,5]{Stefano Liberati}
\author[6]{Massimo Mannarelli}
\affil[1]{Department of Physics, University of Pisa, Largo B. Pontecorvo 3, 56127 Pisa, Italy}
\affil[2]{INFN, Sezione di Pisa, Largo B. Pontecorvo 3, 56127 Pisa, Italy}
\affil[3]{Scuola Internazionale Superiore Studi Avanzati -- SISSA, Via Bonomea 265, 34136 Trieste, Italy}
\affil[4]{Institute for Fundamental Physics of the Universe -- IFPU, Via Beirut 2, 34014 Trieste, Italy}
\affil[5]{INFN, Sezione di Trieste, Via Valerio 2, 34127 Trieste, Italy}
\affil[6]{INFN, Laboratori Nazionali del Gran Sasso, Via G. Acitelli, 22, I-67100 Assergi (AQ), Italy}
\date{\vspace{-5ex}}
\begin{document}
\maketitle

\begin{abstract}
{In this perspective review, we present a concise yet multifaceted overview of the pivotal role played by the the shear viscosity to entropy density ratio  across various physical contexts. After summarizing some of the main aspects of the  bound obtained by  Kovtun, Son and Starinets, we examine potential sources of its violation, exploring the insights these may offer and their connections to fundamental causality conditions. We also review a range of experimental tests conducted in diverse, yet complementary, physical systems, discussing the prospects opened by upcoming measurements.}
\end{abstract}

\noindent {67.10.Jn} {Hydrodynamics in Quantum Fluids}\\
\noindent {04.20.-q} {Einstein equation}\\
\noindent {03.30.+p} {General Relativity}
\newpage

\section{Introduction}
The microscopic description of strongly correlated systems is extremely challenging. When  conventional perturbation theory cannot be employed, the standard approach is to resort to 
numerical techniques. However, they may be structurally limited, as it occurs e.g. in lattice methods, which are hindered by the 
sign problem for Fermi systems and not well-suited to explore real-time processes. One alternative  approach is provided by the seminal concept of AdS/CFT correspondence~\cite{Adams:2012th}. 
This relates gravity on a higher dimensional ``bulk" manifold to a gauge theory on its boundary. Specifically, a gravitational theory in (D+1)-dimensional Anti-de Sitter (AdS) curved spacetime is equivalent to a conformal field theory (CFT) in D dimensions on its boundary and it is, in this sense,  {\it holographic} \cite{Susskind_1995,Kovtun:2003wp}. 
In this holographic duality, new weakly coupled degrees of freedom dynamically emerge from a strongly coupled system. Difficult problems in strongly interacting quantum field theories, such as those relevant to condensed matter~\cite{Hartnoll:2009sz} or quantum chromodynamics\,\cite{Gursoy:2010fj, Kim:2012ey}, can be mapped to more tractable problems in classical gravity. The additional spatial dimension in the gravitational theory effectively encodes the energy scale of the quantum field theory, functioning as a geometric representation of the renormalization group flow. Conversely, transport properties of the boundary theory can be computed as the response of conserved charges to perturbations by external fields. Each conserved quantity in the finite-temperature field theory corresponds to a field in the dual gravitational theory: for instance, the stress-energy tensor maps to the graviton, while conserved charge currents correspond to components of a five-dimensional Maxwell field~\cite{Schafer_2009}.

The gravitational field configuration relevant to field theories at finite temperature $T$ is an AdS$_5$ black hole (BH). BHs can be viewed as thermodynamical objects with an entropy proportional to the cross-sectional area of the event horizon divided by the square of the Planck length \cite{Bekenstein:1973ur} and $T=T_H$, i.e.  {the one set by their Hawking radiation} \cite{Hawking:1975vcx}. The BH fills the AdS {spacetime}  with a bath of gravitational radiation, whose temperature coincides with that of the boundary field theory. The graviton propagation dynamics in the BH background determines the finite-temperature correlations in the stress tensor at the boundary field theory, and therefore the transport coefficients, like the shear viscosity $\eta$ and the bulk viscosity  {$\zeta$,} according to the Green-Kubo equations~\cite{Schafer_2009}. The BH membrane paradigm \cite{Damour:2008ji,membrane} showed that this behavior also extends to non-equilibrium thermodynamics hence to dynamical horizons which behave as viscous friction fluids with negative bulk viscosity. 
In the duality,  a classical BH in AdS spacetime corresponds to a strongly coupled thermal CFT on the boundary at $T=T_H$. The ’t Hooft coupling  $\lambda = g_{\text{YM}}^2 N$ of the $\mathcal{N} = 4$ gauge theory is related to the ratio of the AdS curvature radius $L$ to the string length $l_s$ via $\lambda = (L/l_s)^4$. As a result, the large-scale dynamics of the BH is dual to the hydrodynamics of the thermal gauge theory~\cite{Rangamani:2009xk}.

Clearly, not all strongly correlated systems possess known gravity duals, and most AdS/CFT results pertain to idealized CFTs with a large number of degrees of freedom (colors) $N$ and supersymmetric conditions. Nevertheless, the correspondence provides valuable qualitative insights  {to} build intuition and offer useful diagnostics for the macroscopic behavior of strongly correlated fluids at large distances and long timescales~\cite{Adams:2012th}. This is particularly significant when scaling symmetries and universality play a central role in the underlying physics, as in quantum criticality during phase transitions, the behavior of non-Fermi liquids or strange metals, and universal transport phenomena. Indeed, the holographic duality is particularly useful in the computation of transport coefficients such as viscosity, conductivity, and diffusion constants.

\begin{figure}
\centerline{\includegraphics[width=9.5 cm]{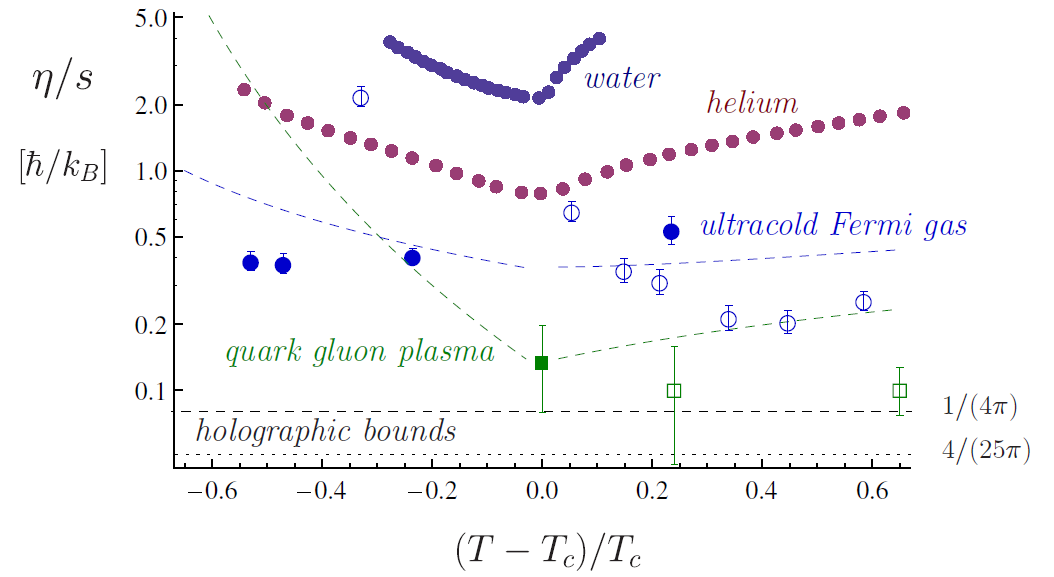}}
\caption{Transport properties of strongly correlated fluids. Shear viscosity $\eta$ to entropy density $s$ ratio vs.~$(T-T_c)/T_c$, with $T_c$ some transition temperature, like e.g.~to superfluidity in ultracold Fermi gases, deconfinement in quantum-chromodynamics  and for liquid-gas transition in water and helium. Dashed curves:  theoretical predictions scaled by overall factors to match the data near $T_c$. Holographic bounds: KSS $1/4\pi$ and Gauss–Bonnet $(16/25) (1/4\pi)$ bounds ($\hbar/k_B$ units). Notice that these systems span 19 orders of magnitude in $T$ and 44 in pressure. From~\cite{Adams:2012th} with included references. (Permission to use this content granted by CC Licence https://creativecommons.org/licenses/by-nc-sa/3.0/).}
\label{fig:1}
\end{figure}

In this context, a prominent role is played by the Kovtun, Son, and Starinets (KSS) \cite{Kovtun:2004de} bound limiting the shear viscosity to entropy ratio. In the limit of infinite coupling,   any gauge theory with an Einstein gravity dual has a ratio of shear-viscosity coefficient  to entropy- {density}
\begin{equation}
\frac{\eta}{s} = \frac{\hbar}{4\pi k_B}    \,,
\label{eq:KSS}
\end{equation}
hence ${\eta}/s = 1/4\pi$ in natural units, which we will use in the following. The KSS bound  is the conjecture that the above $\eta/s$ value is the lower one achievable for any physical system, and that it is  universal and independent of the fine details of the underlying microscopic description, see Fig.~\ref{fig:1}. A number of specific-to-general conditions have been explored under which violations of the KSS bound have been found. Nonetheless, investigating such violations --- or the existence of a universal lower bound different from the KSS value --- is particularly valuable, as it can shed light on our understanding of transport in strongly coupled fluids and, on the gravity side, offer gauge-theoretic interpretations of higher-derivative couplings or new perspectives on causality constraints in gravitational theories. As quantum technologies 2.0 increasingly enable the engineering of transport properties in strongly coupled fluids under highly controlled conditions, these high-level connections between table-top condensed matter analogues, fundamental interactions, and cosmological concepts become both timely and promising for advancing our understanding in these domains.

In this perspective review, we thus first recollect the main derivations of the KSS bound, to form an intuition on its importance. We then identify the possible sources of violation or, vice versa, the extent to which a universal bound may exist. We then review the outcomes of experimental tests of universality designed so far. Finally, we draw the perspective of forthcoming research in this field. 

\section{Discussion of the KSS bound}
Shear viscosity is a measure of a fluid's resistance to shear stress. A good fluid is by definition one with low viscosity, as indeed it can easily flow. However, a fluid with zero viscosity becomes {\it dry}~\cite{Feynman1965}, so in some sense ceases to be a fluid. Entropy is instead a measure of disorder. The very fact that these two quantities can be somehow linked may sound mysterious. For instance, it is well known that  the  maximum entropy is the state of equilibrium in isolated systems. No similar statement can be made for the shear-viscosity  coefficient. Certainly,  both quantities depend on interaction processes,  although their dependence is quite different. Interactions are necessary to reach an equilibrium state, thus to maximize the entropy. At the same time,  interactions tend to hamper diffusion processes, thus  {reducing $\eta$}. For this reason one would expect that with increasing interaction strength, $\eta/s$ may reach a minimum. In physical systems $\eta/s$ approaches a minimum close to  phase transitions, see Fig.~\ref{fig:1}, due to a rapid change of degrees of freedom and interaction strengths. Quite commonly, in the broken  {low-$T$} phase, the effective number of  low-energy degrees of freedom and their interaction strengths increase with  {$T$},  while in the unbroken  {high-$T$} phase,  (quasi)particles diffusion processes become more effective at  {high $T$}. 

A number of apparently very different derivations of the KSS bound have been elaborated. For reference, the shear viscosity coefficient of a dilute gas is $\eta \sim p\,  \ell\, n  $, 
with $p \sim T^{1/2}$ the average momentum, $\ell$ the mean free path and $n$ the number density. From $s \propto n$ at weak coupling and the Heisenberg principle follows that 
$\eta/s\sim p \, \ell \gtrapprox 1$.

This said, we {recall} in brief the holographic proof~\cite{Schafer_2009}. {This hinges on two} key ingredients~\cite{Kovtun:2004de, Son:2007vk}. 
The first  is the Bekenstein--Hawking entropy for 5D AdS--Schwarzschild black brane $S=A/4G_5$,  with $G_5$ the 5D Newton constant. The area of such homogeneous planar horizon $A$ is infinite, but one can suitably define a (finite) area density $a$, normalizing it by the volume $V$ spanned in the non-radial spatial dimensions of the brane, $a=A/V$. This is a dimensionless quantity given by the cube of the ratio between the black brane horizon radius $r_H$ and the AdS curvature radius $L$, $a=(r_H/L)^3$.
Thus the entropy density is  $s=a/(4G_5)$. The second element is that the shear viscosity
can be expressed in terms of the total graviton absorption cross section $\sigma_0$ and Newton constant via the optical theorem as $\eta=\sigma_0/(16\pi G_5)$.
The KSS bound follows from the low energy limit of $\sigma_0$ being equal to the area density $a$, so that $\eta/s=1/4\pi$ {, clearly independent of} $a$ and $G_5$.
Corrections to the infinite coupling limit in ${\cal N} = 4$ supersymmetric Yang-Mills theory lead to {positive contributions to $\eta/s$ \cite{Buchel:2004di}}.
This observation led Kovtun, Son, and Starinets to conjecture that $\eta/s \geq 1/4\pi$. Notably, 
{neither the speed of light nor the gravitational constant are involved in}
the KSS bound, suggesting that its validity could extend beyond specific {relativistic or} gravitational models, even though its original derivation is deeply rooted in gravitational physics.

For the numerical evaluation of  $\eta$, the two most commonly used methods  are the Chapman--Enskog expansion  and the celebrated Green--Kubo formula\,\cite{Green:1954ubq, Kubo:1957mj}. The former solves the Boltzmann equation by applying a systematic integration of moments of the distribution function\,\cite{huang2000statistical, DeGroot:1980dk}. The latter involves finite $T$ Green’s functions of conserved currents and relates the transport coefficients of a slightly non-equilibrium system to real-time correlation functions computed in an equilibrium thermal ensemble. Using the latter method,  it was  shown that for  a Rindler horizon  ${\eta}/s = 1/(4 \pi)$\,\cite{Chirco:2009dc}, 
{thus, explicitly showing the saturation of the KSS bound without employing the initially proposed holographic models.} {This is a crucial aspect, suggesting that it is worth   exploring the  ${\eta}/s$ bound by a wide class of  models having different interactions strengths and $D$ spacetime dimensions.} 

In this context some of us~\cite{LuisaChiofalo:2022ykx} proposed a new approach where the KSS bound was derived within the framework of {analogue gravity, employing} kinetic theory {grounded} {on} geometric considerations.
The basic realization  {of analogue gravity} is that the low-energy excitations (the phonons) of a flowing fluid in Minkowski spacetime can be described in terms of scalar particles embedded in an effective gravitational background. The emerging acoustic metric tensor is determined by the fluid’s properties (see e.g.~\cite{Barcelo:2005fc}).   
{For illustrative purpose,  consider an acoustic spherically symmetric BH analogue, with a radially oriented fluid velocity $\bm v(r)$}  and sound velocity $c_s(r)$. The  sonic horizon is at $r_H$ defined by $c_s(r_H) = v(r_H)$  and the metric reads {
$ds^2 =dt^2 (c_s^2-v^2) \gamma^2 + 2 \gamma^2(1-c_s^2) v dr dt- [(1-c_s^2)\gamma^2 v^2 +1]dr^2 - r^2 d \Omega^2$}
with $\gamma$ the fluid Lorentz factor. {Since the polar and azimuthal coordinates do not play any role, the system has effectively $D=2$.} It can thus be shown \cite{Mannarelli:2021olc} that the horizon develops an outgoing  phonon entropy current $s^\alpha_{\rm ph}$, of purely geometrical origin. This is actually a quite general phenomenon. Considering a  collisionless fluid, the covariant entropy conservation   $\displaystyle( s^\alpha_{\rm ph})_{;\alpha} = 0$ implies that
$\displaystyle\de_\alpha s^\alpha_{\rm ph} = - \Gamma^\mu_{\mu \nu} s_{\rm ph}^\nu $,
 where $\displaystyle \Gamma^\mu_{\mu \nu}$ are the pertinent Christoffel symbols. That is, space-time variations of the entropy flux can be  {derived via} the effective metric.
In the presence of an horizon, the phonon flux can be associated to  fluctuations of the horizon. 
Even in the absence of gravitational field equations, an entropy can be associated to the acoustic horizon via the entanglement of phonon across the horizon, still providing an entropy proportional to the horizon area as in the Bekenstein--Hawking formula.
In this case, any perturbation of the horizon radius determines a variation of the associated entropy, and indeed, the spontaneous phonon emission at the horizon does perturb the horizon itself. Equating the corresponding entropy variation  with that of the phonons calculated using the Christoffel symbols, the {phonon gas temperature} turns out to be~\cite{Mannarelli:2020ebs} 
\begin{equation}
\label{Eq:HawingT}
{T_H} = \frac{1}{2 \pi} \left.\left(\frac{ c_s-|v|}{1-c_s |v| }\right)'\ \right\vert_H\,,
\end{equation}
which coincides with the expression determined by Unruh \cite{Unruh:1980cg} (in the non-relativistic limit) and then by Bili{\'c}\cite{Bilic:1999sq} via a completely different strategy. 
Since the spontaneous phonon emission is an irreversible process, it effectively produces   dissipation: transformation of the fluid bulk kinetic energy in heat. Unlike standard dissipative process, here dissipation   is localized in a region close to the acoustic horizon, where the spontaneous phonon emission takes place.  Let us explicitly see this by following closely the setting presented in~\cite{LuisaChiofalo:2022ykx}. For a planar horizon  {at some fixed $x$ and along the $y$-direction}, a shear perturbation can be  introduced assuming that the velocity is oriented along the negative $x$-axis and its modulus is given by $\displaystyle v = c_s - Cx + Ky $, where $C$ and $K$ are two parameters, with $K \ll C$ to ensure a negligible flow along the $y$ direction. This gives rise to a small tilt of the horizon plane with an angle $\theta \simeq - K/C$ with respect to the $y$-axis. {In what follows we shall take $C$ to be constant and induce a tilt by the sole variation of $K$.
Under these conditions, the leading contribution to the viscous stress tensor is~\cite{LuisaChiofalo:2022ykx}
\begin{equation}
\label{eq:sigmap}
\sigma_{ik} = \eta  (\delta_{iy}\delta_{kx} \partial_i v_k+\delta_{ix}\delta_{ky} \partial_k v_i) + \zeta \delta_{ix}\delta_{kx}{\bm \nabla} \cdot {\bm v}\,,
\end{equation}
where the viscosity coefficients can be determined}  equating the viscous stress tensor of the fluid to the energy-momentum tensor of phonons. 
Now, the phonons are emitted orthogonally to the horizon (see \cite{Mannarelli:2021olc}) with a Planckian distribution at $T_H=C/2\pi$. Such flux is associated to an energy density and pressure, the latter being both along the $x$-direction, $P_{\text{ph} \, x}$, as well as along the, transverse, $y$-direction, $P^y_{\text{ph} \, x}$. As said, the spontaneous emission of phonons at the horizon leads to an irreversible increase in heat, at the expense of the bulk fluid's kinetic energy.   Equating the viscous stress
with the corresponding phonon pressure terms, i.e.~$\sigma_{xx}= P_{\text{ph} \, x}$, $\sigma_{xy} = P^y_{\text{ph} \, x}$ and  using the Gibbs--Duhem relation
$\displaystyle{ \epsilon_\text{ph} = -P_{\text{ph} \,x} + Ts_\text{ph} }$, one gets $\displaystyle {\zeta}/{ s_\text{ph}} ={\eta}/{ s_\text{ph}} = {1}/({4 \pi})$.
{In fact, the} phonon and the horizon entropies are distinct quantities. The former is the thermodynamic entropy of the (phonon) gas, the latter is instead associated to the horizon area. The variations of the two entropies are clearly  connected: a variation of the horizon's area modifies the source of the spontaneously emitted phonons. An independent derivation, more in the spirit of the BH membrane paradigm, can be found in~\cite{LuisaChiofalo:2022ykx}, where the shear viscosity of the deformed/stretched horizon is directly related to the variation of  its cross-sectional area element which, due to the spatial partition of the vacuum state, yields an entanglement entropy~\cite{Finazzi:2013sqa} in agreement with the Bekenstein--Hawking area law
\cite{Bekenstein:1973ur}.

\section{{Violations and alternative lower bounds}}   
The different KSS bound derivations suggest a high degree of generality~\cite{Kovtun:2003wp,Policastro_2002,Herzog_2002}. All gauge theories at infinite coupling  {holographically} dual to classical, two-derivative (Einstein) gravity, saturate the bound irrespective of matter content, amount of supersymmetry and presence of conformality~\cite{Buchel, Cremonini:2011iq}. However, deviations from conformality in theories with no gravity dual  may induce violations~\cite{Trabucco:2025duz}. It was also demonstrated~\cite{ChircoElingLiberati_2011} that in the case of a gravity-Rindler fluid duality, the dual metric solution at lowest-orders in the non-relativistic expansion is a solution to General Relativity and to any higher curvature theory of gravity, the  transport coefficients being changed only in their second order form.

A change in the stress energy tensor embodying $\eta$ on the  CFT side is dual to the metric on the gravity side, notwithstanding that  {$s$} can be conceptually connected to time-arrow properties. The possible sources of violation can be classified in higher-derivative corrections on the gravity side, corresponding to finite coupling $\lambda$ and degrees of freedom $N$ on the CFT side, and violations of causality conditions\,\cite{Cremonini:2011iq}. {Some of these violations can lead to alternative lower bounds than KSS. In any event,} relevant implications arise, as discussed in the {perspectives} section.\\
\textit{Higher-curvature corrections.} Let us start with the case of higher-curvature corrections to the gravitational action, generally expressed in the form  
\begin{equation}
S=\frac{1}{16\pi G_N}\int d^Dx \sqrt{-g}[R-\Lambda-F^2+\Sigma_n(\alpha')^{n-1}R^n],
\end{equation}
with $R^n$ any contraction involving $n$ Riemann tensors suppressed with powers of $\alpha'=l_s^2$ and $l_s$ the string length. 
Here, specific cases can be constructed in which the bound is either reinforced or violated.  For example,  for quartic curvature corrections $\sim \alpha'^{3}R^4$ the bound is reinforced.  Specifically in type IIB supergravity on $AdS_5\times S^5$, dual to $\mathcal{N}=4$ SYM theory, , the bound reads $4\pi\eta/s=1+15\zeta(3)/ \lambda^{3/2}$. 
The bound is violated for e.g. quadratic corrections with positive coupling $\lambda_{GB}$ of a Gauss-Bonnet type: $S=(16\pi G_N)^{-1}\int d^5x \sqrt{-g}[R-2\Lambda+\lambda_{GB}/2 L^2 (R^2-4R_{\mu\nu}R^{\mu\nu}+R_{\mu\nu\rho\sigma}R^{\mu\nu\rho\sigma}]$  and  $4\pi\eta/s=(1- 4\lambda_{GB})$. 

Kats and Petrov~\cite{Kats_2009} have analyzed the more general case of curvature squared corrections in the action $S=(16\pi G_N)^{-1}\int d^Dx \sqrt{-g}[R/(2\kappa)-\Lambda+c_1R^2+c_2R_{\mu\nu}R^{\mu\nu}+c_3R_{\mu\nu\rho\sigma}R^{\mu\nu\rho\sigma}] $   with $c_i$ arbitrary small coefficients and the negative cosmological constant $\Lambda$ creating an AdS space with radius $L^2=(D-1)(D-2)/(2\kappa(-\Lambda))$. After using the real time AdS/CFT correspondence~\cite{Herzog_2003}, they find that only $c_3$ affects the $\eta/s$ ratio ($c_1$ and $c_2$ affecting $\eta$ and $s$ individually) as $4\pi\eta/s=(1-4(D-4)(D-1)c_3\kappa/L^2)$, {so that the KSS bound is} satisfied  {in dimensions $D\ge 5$,}   only if $c_3<0$.
{In these models no correction that violates  the KSS bound arises for $D=4$.  Regarding $c_3$, } while consistently restricting {its}  sign is not generally possible in low-energy effective field theories,  {in the limit of large $N$ and constant ‘t Hooft coupling}, the $c_3$ coefficient  {becomes} related to the CFT central charge $\mathfrak c$, through $c_3(\kappa/L^2)\simeq (c-a)/(16c)$, yielding $4\pi\eta/s=a/c$  {up to order} $1/N^2$.  For the $\mathcal{N}=2 Sp(N)$ gauge theory with $4$ fundamental and $1$ antisymmetric traceless hypermultiplet,  leading to $4\pi\eta/s=(1-1/(2N))$~\cite{Kats_2009} and thus a violation. 

To gain a more general understanding, it is to be noted that in holography the coefficients of higher-derivative terms  can be related to the central charges $\mathfrak a$ and $ \mathfrak c$ entering the trace of the stress-energy tensor\cite{Cremonini:2011iq}. {Let us write} the latter as $\langle T^\mu_\mu\rangle ={\mathfrak c}/({16\pi^2})(Weyl)^2-{\mathfrak a}/({16\pi^2}) (Euler)$, with $(Weyl)^2=R^2_{\mu\nu\rho\sigma}-2 R^2_{\mu\nu}+R^2/3$ and $Euler= R^2_{\mu\nu\rho\sigma}-4 R^2_{\mu\nu}+R^2${: it can thus} be shown that for an action of the form $(2\kappa^{2})^{-1}[R+12g^2+\alpha R+\beta R^2_{\mu\nu}+\gamma R^2_{\mu\nu\rho\sigma+…}]$, the couplings $\alpha,\beta,\gamma,…$ of the four-derivative terms in the Lagrangian,  geometrical quantities, can be connected to $\mathfrak a$ and $\mathfrak c$. For example, it can be derived that $({\mathfrak c}-{\mathfrak a})/{\mathfrak a}=\gamma/L^2$, highlighting that a difference in the central charges is sensitive to the $R^2_{\mu\nu\rho\sigma}$ term. {As a result,} finite  $\lambda$ (inverse ‘t Hooft) coupling effects are associated with $R^4$ terms and preserve the KSS bound, while finite $N$ effects are associated with $R^2$ terms and can violate the KSS bound (if ${\mathfrak c}>{\mathfrak a}$), consistently with the observation that $({\mathfrak c}-{\mathfrak a})\sim N$.

The bound violation due to higher-derivative corrections can be enhanced by the inclusion of a chemical potential, that in the gauge/gravity duality context can be mimicked by an R-charge in the presence of an R-symmetry. For four-derivative corrections to gravity in five dimensions, coupled to a $U(1)$ gauge field with charge $q$ and a negative cosmological constant, it can be shown~\cite{Myers_2009} that $4\pi\eta/s=1-8c_1+4(c_1+6c_2)q^2/r_0^6$, with $r_0$ the horizon radius and $c_{1,2}$ the coefficients of $R_{\mu\nu\rho\sigma}R^{\mu\nu\rho\sigma}$ and $R_{\mu\nu\rho\sigma}F^{\mu\nu}F^{\rho\sigma}$, respectively. For $\mathcal{N}=2$ gauged supergravity, the  {violation range} can be summarized as $[1-3({\mathfrak c}-{\mathfrak a})/{\mathfrak a}]\leq 4\pi\eta/s\leq [1-({\mathfrak c}-{\mathfrak a})/{\mathfrak a}]$ and estimated to be small.   
To conclude the analysis of the higher-derivative sources of violation, Cremonini~\cite{Cremonini:2011iq} highlights that only terms explicitly dependent on the Riemann tensor affect $\eta/s${, reminding} the  Wald’s entropy formula for BHs with higher derivatives, the remarkable simplicity of the ratio further reinforcing the inference of forms of universality.

\textit{Causality conditions.} The analysis of the Gauss--Bonnet (GB)  model is instructive to explore how much {the KSS bound on} $\eta/s$ can be lowered and, at the same time, the connection of this outcome  with causality conditions\,\cite{Brigante:2008gz}. In GB the couplings of the higher-derivative terms are finite (i.e.~not perturbatively small), BH solutions are known for arbitrarily large couplings, though $\lambda_{GB}\leq 1/4$ to ensure a vacuum AdS solution and a CFT dual~\cite{Cremonini:2011iq}. To preserve causality,
$\lambda_{GB}<9/100$,  this  excludes  microcausality violation in the CFT\,\cite{Brigante:2008gz}, leading to $16/25\leq 4\pi\eta/s$. Even lower bounds have been found, $7/16\leq 4\pi\eta/s$, for AdS$_7$ GB~\cite{deBoer_2010} and  $0.41\leq 4\pi\eta/s$ for  quasi-topological gravity, which includes curvature-cubed interactions~\cite{Myers_2010}. To identify general connections between causality conditions and KSS bound violation, it is important to note~\cite{Cremonini:2011iq} that transport coefficients like $\eta$ depend on the  infrared (IR) dynamics, while causality constraints arise from the ultraviolet (UV) behavior associated to high-energy, short-distance physics. Therefore, any observed connection between $\eta/s$ and causality must be capable of crossing over between IR ($\omega,|k|\ll \min(T,\mu)$) and UV $\omega,|k|\gg \max(T,\mu)$ {opposite regimes}. The Gauss--Bonnet plasma is one such example, being a conformal theory. This {extremely reinforces the interest} to explore situations {where} the bound is violated  {due to the} occurrence of phase transitions or other qualitative changes across energy scales. For instance,  in~\cite{Buchel_2010}, a GB model with a superfluid phase transition was constructed, where above $T_c$ the gravitational theory is simply Einstein gravity and saturates the bound, while  {for $T<T_c$} not even a lower bound exists, though no superluminal causality-violating modes appear. From this specific example a more general behavior can be conjectured: whenever, for whichever reason, the IR transport and UV causality properties decouple, microcausality does not necessarily set a lower bound for $\eta/s$. For example, non-trivial temperature flows connecting IR and UV regimes, which may correspond to different behaviors of the gravitational background~\cite{Cremonini:2011iq}.

\begin{figure}
\centerline{\includegraphics[width=8cm]{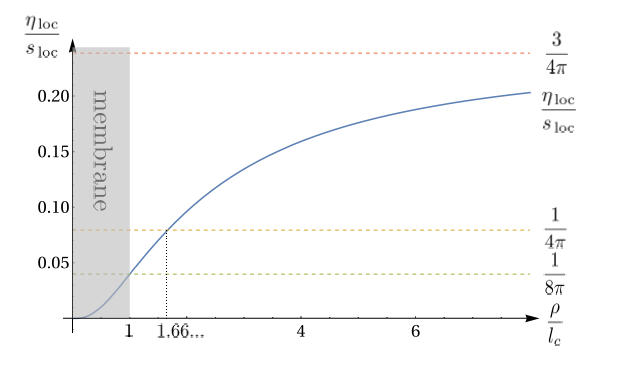}}
\caption{Local $\eta/s$ ratio for massless fields with spins 1/2 and 1 (blue solid line)  {vs.} the ray coordinate $\rho$. Shaded gray region: area between {extended} and true horizons, corresponding to the membrane thickness $0\leq\rho\leq l_c$. From ~\cite{Lapygin_2025}. (Permission to use this content granted by CC Licence https://creativecommons.org/licenses/by/4.0/).}
\label{fig:2}
\end{figure}

\textit{{Accelerated frames.}} 
We conclude with the more recent findings from Lapygin et al.~\cite{Lapygin_2025}, generalizing the derivation of~\cite{Chirco:2010xx} for the dissipative properties of different quantum fluids living above the Rindler membrane, to the case of massless fields with spins 1/2 and 1. They  show that the viscosity in the accelerated frame of the Minkowski vacuum state always satisfies the KSS bound. However, at a given distance from the Rindler horizon this happens only on average, while it is violated near the membrane, becoming half of the usual KSS bound, i.e.~$\eta/s=1/(8\pi)$, at the outer boundary of the membrane on the stretched horizon (see Fig.~\ref{fig:2}). 
 
\section{Experimental tests}   
Experimental {measurements} of  $\eta/s$  have been conducted in various fluids, including water, liquid Helium $^4$He, ultracold atoms, quark gluon plasma and charge transport in semiconductors (Fig.~\ref{fig:1}).\\ 
\noindent\textit{Liquid Helium}. In liquid $^4$He, $\eta$ can be measured from capillary flow, angular velocities of rotating disks and damping of oscillating spheres or plates~\cite{Schafer_2009}. The minimum $\eta$ at normal pressure occurs just below the $\lambda$ point with $\eta\simeq 1.2\cdot 10^{-5}$ Poise~\cite{Woods}. The minimum $\eta/s\simeq 0.8$ is instead located at higher $T$ close to the liquid-gas transition.\\  
\noindent\textit{Ultracold gases}. Ultracold quantum gases offer a versatile playground to probe concepts in curved spacetime. {The physical reality of phonon Hawking emission of sonic BHs has  been predicted numerically\, \cite{Carusotto} and experimentally realized} with atomic Bose--Einstein condensates\, \cite{Steinhauer,MunozdeNova:2018fxv}. {Transport coefficients can generally} be extracted from the free expansion of a deformed trap, e.g. measuring the elliptic flow, the damping of collective oscillations, sound propagation, and expansion out of rotating trap. {Through the tuning of the atom interaction strength by a variety of means, including the preeminent Fano--Feshbach resonance~\cite{Schafer_2009}, various regimes can be explored}. The entropy can be determined in a nearly model-independent way by indirectly connecting it to the atomic cloud size. The total energy  can be inferred from the cloud size in the unitary, strongly correlated, thus universal regime, then the entropy vs.~energy can be determined  via the Fano--Feshbach mechanism, i.e.~through an adiabatic magnetic field sweep that transfers the gas to a weakly interacting regime, where thermodynamic properties are well understood~\cite{Luo_2007}. Seminal experiments with atomic Fermi gases~\cite{Thomas_2009} show that $\eta/s$ is close to 0.5 in proximity of  $T_c$.

To infer $\eta$ in ultracold atoms, it is especially useful to exploit the scissor-mode~\cite{Gu_ry_Odelin_1999} that sets in due to the different moment of inertia of  the normal and superfluid components. The obtained  values of $\eta/s$ are close to the $1/4\pi$ bound~\cite{Clancy_2007,Thomas_2009}.  This notwithstanding, a few challenges should still be faced before considering these tests as definitive: experimental setups should be improved to assess that shear viscosity is the only source of dissipation, that transport behavior is surely hydrodynamic, and that cloud inhomogeneities are not relevant. More recent exciting advances {regard the measurement of entropy and density response functions through direct thermography~\cite{Zwierlein_2024} and in flat trapping~\cite{Hadzi_2021}: if exploited in combination with interaction-strength control, they} could provide a very promising testbed with unprecedented accuracy.\\ 
\noindent\textit{Quark-gluon plasma}. Ultracold Fermi gases are characterized by $10^{-6}-10^{-9}$ K temperature and $10^{-7}-10^{-11}$ Pa pressure ranges. At the opposite extreme of, say, $10^{12}$ K and $10^{35}$ Pa is the quark-gluon plasma (QGP). Studies involving the QGP created in relativistic heavy ion collisions have been conducted at AGS (Brookhaven), SPS (CERN), RHIC (Brookhaven). Experiments performed at   RHIC~\cite{Romatschke:2007mq, Song_2012, Xu:2007jv, Ferini:2008he} use heavy-ion collisions at energies of  hundreds of  GeV  per nucleon.
Though difficult to analyze~\cite{Romatschke:2007mq, Song_2012, Xu:2007jv, Ferini:2008he}, the spectra of produced particles indicate that conditions of local equilibration are reasonably met and the measurement of anisotropic flow expansion can be used to extract the shear-to-entropy ratio\,\cite{Schenke:2010rr}. From the data\,\cite{STAR:2005gfr,PHOBOS:2004zne, BRAHMS:2004adc, PHENIX:2004vcz}, a conservative $\eta/s<0.4$ can be inferred, though a best fit {provides a lower} $\eta/s\sim 0.1$ near $T_c$ (see Fig.~\ref{fig:1}), i.e.~very close to  the KSS bound.\\ 
\noindent\textit{Solid-state platforms}. We close this account of experimental tests with the emergent case of solid-state platforms, where transport properties of strongly correlated electron fluids can be probed. Though clear tests are not yet available, interesting predictions are. Graphene attracts special attention, due to its massless relativistic quasiparticles dispersion and their high mobility {at} relatively high $T$: in fact, a system conceptually not far from ultrarelativistic QGP. In ~\cite{Muller_2009}, it is demonstrated by quantum kinetic theory methods that quantum criticality in undoped graphene drives  {a smaller $\eta/s$} than in many other correlated quantum liquids and close to the QGP prediction. 

Similar ideas could be played in two-dimensional (2D) and 3D quadratic-band touching semimetals~\cite{Kim_2021}. In~\cite{Ge_2020}, a translationally and rotationally invariant 2D strongly correlated solvable model is investigated, {in the form of} coupled Sachdev-Ye-Kitaev  islands. This model is believed to have a gravity dual and exhibits an intermediate-temperature incoherent metal regime {with broken-down quasiparticle picture, resulting in a, strong KSS bound violation for} a robust temperature range. In ~\cite{Link_2018}, anisotropy effects are explored in the hydrodynamic transport of 2D interacting electronic systems with merging Dirac points at charge neutrality. Due to Dirac- and Newtonian-like excitation dispersion along orthogonal crystallographic directions, the electrical conductivity is either metallic or insulating, respectively: at $T=0$, one of the shear viscosity components vanishes, leading to a KSS bound violation.  Practical solid-state model implementations include organic conductors like $\alpha(BEDT-TTF)_2I_3$ under pressure, $5/3$ $TiO_2/VO_2$ supercell heterostructures, surface modes of topological crystalline insulators with unpinned surface Dirac cones, and quadratic double Weyl fermions~\cite{Muller_2009}. 

While measuring $\eta/s$ in solid-state platforms poses additional challenges due to unknown effects from boundary roughness, there are significant experimental advances. For instance, in\,\cite{Keser_2021} conventional modulation-doped GaAs/Al heterostructures could be equipped with  smooth sidewalls to impose perfect slip conditions, allowing to isolate intrinsic viscous effects and observing a transition from ballistic to hydrodynamic flow driven by temperature and magnetic field. Under these conditions, a strong density dependence of the viscosity has been found, that is not captured by standard theories and that requires improved electron-fluid models.

\section{Perspectives}   
The KSS bound serves as an elegant tool to 
{probe the physics of strongly correlated fluids as well as gravitational systems. It }
holographically connects a broad class of $D$-dimensional gauge theories with gravitational theories in a $(D+1)$-dimensional AdS spacetime, offering valuable insights, particularly when scaling symmetries and universality play a pivotal role. Investigating possible violations {or universal lower bounds than KSS} value{, originated by either higher-curvature corrections or causality conditions or else reference-frames effects,} one can extract important lessons about the microscopic dynamics of fluids both at and out of equilibrium, especially near quantum phase transitions. These insights may also suggest gauge-theoretic interpretations of higher-derivative gravitational couplings or offer new perspectives on causality constraints.

Since transport coefficients probe the infrared, long-distance dynamics, while causality constraints pertain to the ultraviolet, short-distance behavior, it is not surprising that a single system may exhibit violations under different regimes, each corresponding to distinct features of the gravitational background. Therefore, exploring scenarios in which the KSS bound is violated due to phase transitions or other qualitative changes across energy scales can be extremely valuable~\cite{Cremonini:2011iq}. From this perspective, if a universal lower bound does exist, its implications could open up entirely new and unexplored landscapes in the microphysics of strongly coupled systems.

While these ideas have been explored for some time, they have now become particularly timely due to  {their possible concrete experimental testing} across nearly 19 orders of magnitude in temperature: from high-energy collider experiments to table-top setups enabled by current quantum technology platforms. The increasing precision and controllability of these platforms is driving the development of ingenious techniques that allow for direct access to relevant observables, thereby reducing interpretation uncertainties. At the same time, they offer a unique opportunity to tune a single system across a wide range of energy scales, where the most intriguing phenomena are likely to emerge. We believe this is an exceptionally opportune moment to pursue research in this direction.\\\\\\

\noindent{\textbf{Acknowledgments}}\\
\noindent This perspective review is nurtured by a collaborative and interdisciplinary environment we have established along the years, which bright students and younger researchers have extraordinarily contributed to: we wish to thank in particular Silvia Trabucco, Alessia Biondi, Chiara Coviello and Luca Lepori.   


\end{document}